\documentclass[12pt]{article}
\usepackage{amsmath,amssymb,amsfonts,amsthm}

\title{Algebraic and group treatments to nonlinear displaced number states and their nonclassicality features}

\author{N Asili Firouzabadi$^{1}$, M K Tavassoly$^{1,2}$ and M J Faghihi$^{3,*}$ \\
 \footnotesize{$^1$ Atomic and Molecular Group, Faculty of Physics, Yazd University, Yazd, Iran} \\
 \footnotesize{$^2$ The Laboratory of Quantum Information Processing, Yazd University, Yazd, Iran} \\
 \footnotesize{$^3$ Physics and Photonics Department, Graduate University of Advanced Technology, Mahan, Kerman, Iran} \\
 \footnotesize{$^*$ E-mail: mj.faghihi@kgut.ac.ir}}

\begin{document}
\maketitle

 \newcommand{\norm}[1]{\left\Vert#1\right\Vert}
 \newcommand{\abs}[1]{\left\vert#1\right\vert}
 \newcommand{\set}[1]{\left\{#1\right\}}
 \newcommand{\R}{\mathbb R}
 \newcommand{\I}{\mathbb{I}}
 \newcommand{\C}{\mathbb C}
 \newcommand{\eps}{\varepsilon}
 \newcommand{\To}{\longrightarrow}
 \newcommand{\BX}{\mathbf{B}(X)}
 \newcommand{\HH}{\mathfrak{H}}
 \newcommand{\A}{\mathcal{A}}
 \newcommand{\D}{\mathcal{D}}
 \newcommand{\N}{\mathcal{N}}
 \newcommand{\x}{\mathcal{x}}
 \newcommand{\p}{\mathcal{p}}
 \newcommand{\la}{\lambda}
 \newcommand{\af}{a^{ }_F}
 \newcommand{\afd}{a^\dag_F}
 \newcommand{\afy}{a^{ }_{F^{-1}}}
 \newcommand{\afdy}{a^\dag_{F^{-1}}}
 \newcommand{\fn}{\phi^{ }_n}
 \newcommand{\HD}{\hat{\mathcal{H}}}
 \newcommand{\HDD}{\mathcal{H}}

 \begin{abstract}
Recently, nonlinear displaced number states (NDNSs) have been \emph{manually} introduced, in which the deformation function $f(n)$ has been artificially added to the well-known displaced number states (DNSs). In this paper, after expressing enough physical motivation of our procedure, four distinct classes of NDNSs are presented by applying algebraic and group treatments. To achieve this purpose, by considering the DNSs and recalling the nonlinear coherent states formalism, the NDNSs are logically defined through an algebraic consideration. In addition, by using a particular class of Gilmore-Perelomov-type of $SU(1,1)$ and a class of $SU(2)$ coherent states, the NDNSs are introduced via group theoretical approach. Then, in order to examine the nonclassical behaviour of these states, sub-Poissonian statistics by evaluating Mandel parameter and Wigner quasi-probability distribution function associated with the obtained NDNSs are discussed, in detail.
 \end{abstract}



 \section{Introduction}
Nowadays, nonclassical states of the radiation field have obtained a great deal of attention in various fields of research, such as quantum optics, quantum cryptography and quantum communication \cite{Klauder1985,Schumacher1996,Kempe1999,Bennett1999,Ali2000}. These states may be generated through the conditional measurement techniques or the atom-field interactions in cavity QED \cite{Meekhof1996,Zou2002,Deleglise2008}, and also may be revealed, for instance, in the Jaynes-Cummings model \cite{Faghihi.Tavassoly2013,Faghihi.Tavassoly2013a,Faghihi2014,Faghihi2014a,Hekmatara2014,Baghshahi.Tavassoly2014} and in the field of nonlinear coherent states \cite{MatosFilho.Vogel1996,Manko.etal1997,Tavassoly2006,Tavassoly.Parsaiean2007,Tavassoly2008,Honarasa.etal2009,Honarasa2009} that are naturally arisen from the canonical (standard) coherent states. \\
The standard coherent state defined by $\vert\alpha\rangle = \mathrm{e}^{-\vert\alpha\vert^{2}/2}\sum_{n=0}^{\infty} \frac{\alpha^{n}}{\sqrt{n!}}\vert n\rangle$ is quantum state that describes the radiation field which are known as displaced vacuum states, $|\alpha\rangle = \hat{D}(\alpha) |0 \rangle$, where $ \hat{D}(\alpha) = \exp \left( \alpha \hat{a}^{\dagger}-\alpha^{*} \hat{a} \right) $ is the well-known displacement operator in which $\hat{a}$ and $\hat{a}^{\dag}$ are the bosonic annihilation and creation operators, respectively.
Considering this idea, regarding the construction of coherent state, the displaced number states (DNSs) have been introduced by acting the displaced operator on the number state $ | n \rangle $ which are defined by $ | n,\alpha \rangle = \hat{D}(\alpha) | n \rangle$ \cite{Oliveira1990}:
 \begin{eqnarray}\label{DNSs-Oliveira}
\hspace{-1cm}|\alpha ,n\rangle = \mathrm{e}^{-\frac{\vert\alpha \vert^{2}}{2}} \times
 \left \{ {\begin{array}{*{20}{c}}
{\mathcal{N}_{1}\sum_{m=0}^{\infty}\sqrt{\frac{m!}{n!}}(-\alpha^{*})^{n-m}
L^{n-m}_{m}(\vert\alpha\vert^{2})  \vert m\rangle},\;\;\;\;\;\;     m\leq n, \\
\\
\mathcal{N}_{2}\sum_{m=0}^{\infty}\sqrt{\frac{n!}{m!}}\alpha^{m-n}
L^{m-n}_{n}(\vert\alpha\vert^{2})\vert m\rangle,\;\;\;\;\;\;    m\geq n.
\end{array}}\right.
\end{eqnarray}
 It has been shown that, DNSs indicate several interesting nonclassicality features such as unusual oscillations in the photon number distribution interpreting as the interference in the phase space \cite{Schleich1987}. \\
On the other hand, nonlinear coherent states, which are known as a natural generalization of canonical coherent states (corresponding to simple harmonic oscillator) to $f$-deformed ones (associated with nonlinear oscillators) \cite{MatosFilho.Vogel1996,Manko.etal1997}, can be considered as suitable candidates from which nonclassical light comes out \cite{Tavassoly2010,Honarasa2011a,Safaeian.Tavassoly2011,Faghihi.Tavassoly2011,Piroozi.Tavassoly2012}. It is worthwhile to mention that, there exist many generalized coherent states categorizing in this special class of quantum states, which exhibit the nonclassicality features of light, i.e., `nonclassical' light \cite{Ali2004,Roknizadeh.Tavassoly2004,Roknizadeh.Tavassoly2005}. \\
Based on the above explanations, regarding the DNSs as well as the nonlinear coherent states, one may motivate to establish a direct connection between DNSs and nonlinear coherent states, which is led to the concept of `nonlinear displaced number states' (NDNSs). This idea has recently been introduced by de Oliveira {\it et al} \cite{Oliveira2005}. In this attempt, in analogy between the coefficients of the DNSs and the nonlinear coherent states (defined by Man'ko {\it et al} \cite{Manko.etal1997}), the authors have \emph{manually} taken the nonlinearity function $f(n)$ into account in the DNSs which has been led them to the construction of $f$-deformed (nonlinear) DNSs, given by
\begin{eqnarray}\label{NDNSs-Oliveira}
\hspace{-1cm}|\alpha , f,n\rangle = \mathrm{e}^{-\frac{\vert\alpha \vert^{2}}{2}} \times
 \left \{ {\begin{array}{*{20}{c}}
{N_{1}\sum_{m=0}^{\infty}\sqrt{\frac{m!}{n!}}\frac{1}{[f(m)]!}(-\alpha^{*})^{n-m}
L^{n-m}_{m}(\vert\alpha\vert^{2})  \vert m\rangle},\;\;\;\;\;\;     m\leq n, \\
\\
N_{2}\sum_{m=0}^{\infty}\sqrt{\frac{n!}{m!}}\frac{1}{[f(m)]!}\alpha^{m-n}
L^{m-n}_{n}(\vert\alpha\vert^{2})\vert m\rangle,\;\;\;\;\;\;    m\geq n.
\end{array}}\right.
\end{eqnarray}
It seems that, the construction of NDNSs in \cite{Oliveira2005}, in our opinion, is so artificial. In this paper, by modifying the definition of NDNSs in \cite{Oliveira2005}, we intend to outline a logical formalism from which NDNSs are reasonably constructed. For this purpose, by recalling the nonlinear coherent states approach together with the displaced operator, an algebraic method by which the NDNSs are introduced, is presented. In addition, by using a particular class of Gilmore-Perelomov-type of $SU(1,1)$ and a class of $SU(2)$ coherent states, the NDNSs are defined via group theoretical approach. Then, in each case, some of the well-known nonclassicality features are numerically evaluated. \\
The plan of this paper is as follows: In the next section, the NDNS is algebraically introduced. In section 3, by considering two particular classes of coherent states, the NDNS is defined via group approach. Section 4 deals with studying the nonclassicality signs of the obtained NDNSs through the Mandel parameter as well as the Wigner quasi-distribution function. Finally, section 5 contains a summary and concluding remarks.
 \section{Nonlinear displaced number states:  Algebraic approach}
This section is devoted to the construction of the NDNSs via algebraic method. To reach this goal, it is necessary to introduce the generalized displacement operators $\hat{D}_{f}(\alpha)$ by joining the nonlinear coherent state method and the standard displaced operator. So, the generalized displaced operator reads as $ \hat{D}_{f}(\alpha) = \exp \left( \alpha \hat{A}^{\dag} - \alpha^{*} \hat{A} \right) $, in which $\hat{A} = \hat{a} f(\hat{n})$ and $\hat{A}^{\dag} = f(\hat{n}) \hat{a}^{\dag}$ represent the nonlinear ($f$-deformed) annihilation and creation operators, respectively \cite{MatosFilho.Vogel1996,Manko.etal1997}.
Now, the following communication relations are obviously satisfied:
\begin{eqnarray}\label{vrrd2}
 \left[\hat{A}, \hat{A}^{\dag}\right] = (\hat{n}+1)f^2(\hat{n}+1)-\hat{n} f^2(\hat{n}), \hspace{0.7cm}
 \left[\hat{A},\hat{n}\right] = \hat{A}, \hspace{0.7cm} \left[\hat{A}^{\dag},\hat{n}\right]=-\hat{A}^{\dag},
 \end{eqnarray}
 where $f(\hat{n})$ is a Hermitian operator-valued function which depends on the number operator. The relation (\ref{vrrd2}) clearly shows that $ \hat{D}_{f}(\alpha) $ cannot be generally separated because of the commutation relation of $\hat{A}$ and $\hat{A}^{\dag}$ is a complicated operator. In order to dispel this problem and to be able to act the generalized displaced operator on the number state, Roy and Roy  \cite{Roy2000} gave a proposition and defined two new auxiliary operators as follows:
 \begin{eqnarray}\label{Roy2000}
\hat{B} = \hat{a}\frac{1}{f(\hat{n})}, \hspace{0.7cm} \hat{B}^{\dag} = \frac{1}{f(\hat{n})} \hat{a}^{\dag}, \hspace{0.7cm}
\left[\hat{A}, \hat{B}^{\dag}\right] = \left[\hat{B}, \hat{A}^{\dag}\right] = \hat{I}.
 \end{eqnarray}
It is noticeable to state that, the above proposition has been established in a general mathematical framework by Ali {\it et al} in \cite{Ali2004}.
As a consequence of the latter relation, it may be observed that, by considering a special composition of the operators $\hat{A}$ and $\hat{B}$,
the generators $ \left\{ \hat{A}, \hat{B}^{\dag}, \hat{B}^{\dag} \hat{A}, \hat{I} \right\} $ and also $ \left\{\hat{B}, \hat{A}^{\dag}, \hat{A}^{\dag} \hat{B}, \hat{I} \right\}$ constitute the commutation relations of the Weyl-Heisenberg Lie algebra and the following relations clearly hold \cite{Roy2000,Roknizadeh.Tavassoly2004}:
 \begin{eqnarray}\label{ABRelations}
\hat{B}^{\dag} \hat{A} | n \rangle = n | n \rangle = \hat{A}^{\dag} \hat{B} | n \rangle, \hspace{0.7cm}
\left[\hat{A}, \hat{B}^{\dag} \hat{A} \right] = \hat{A}, \hspace{0.7cm}
\left[\hat{B}^{\dag}, \hat{B}^{\dag} \hat{A} \right] = - \hat{B}^{\dag}.
 \end{eqnarray}
As a result, two generalized displacement operators can be defined which are given by
 \begin{eqnarray}\label{TwoCRelations1}
D_{f}^{'}(\alpha) = \exp \left( \alpha \hat{A}^{\dag} - \alpha^{*} \hat{B} \right) =  \mathrm{e}^{- \frac{|\alpha|^{2}}{2}}  \mathrm{e}^{ \alpha \hat{A}^{\dag}}  \mathrm{e}^{- \alpha^{*} \hat{B}},
 \end{eqnarray}
 \begin{eqnarray}\label{TwoCRelations2}
D_{f}^{''}(\alpha) = \exp \left( \alpha \hat{B}^{\dag} - \alpha^{*} \hat{A} \right)  =  \mathrm{e}^{- \frac{|\alpha|^{2}}{2}}  \mathrm{e}^{ \alpha \hat{B}^{\dag}}  \mathrm{e}^{- \alpha^{*} \hat{A}},
 \end{eqnarray}
in which we have used the Baker-Campbell-Hausdorff formula. Now, by the action of two distinct {\it displacement-type} or {\it generalized displacement} operators defined in (\ref{TwoCRelations1}) and (\ref{TwoCRelations2}) on the number state, the NDNSs are introduced in the following ways:
\begin{eqnarray}\label{NDNS1}
| \alpha ,f , n \rangle^{'} = \hat{D}_{f}^{'}(\alpha) | n \rangle,
\end{eqnarray}
\begin{eqnarray}\label{NDNS2}
| \alpha ,f , n \rangle^{''} = \hat{D}_{f}(\alpha) | n \rangle.
\end{eqnarray}
By substituting (\ref{TwoCRelations1}) into relations (\ref{NDNS1}) and after some lengthy but straightforward manipulations, the explicit form of the NDNSs is given by
\begin{eqnarray}\label{NDNS1-Expansion}
\hspace{-2cm}| \alpha ,f , n \rangle^{'} = \mathrm{e}^{- \frac{|\alpha|^{2}}{2}} \times
\left\{
\begin{array}{c}
\mathcal{N}^{'}_{1}\sum_{m=0}^{\infty}\sqrt{\frac{m!}{n!}}\frac{[f(m)]!}{[f(n)]!}
(-\alpha^{*})^{n-m}L^{n-m}_{m}(\vert\alpha\vert^{2})   \vert m\rangle,\;\;\;\;\;\;       m\leq n, \\
\\
\mathcal{N}^{'}_{2}\sum_{m=0}^{\infty}\sqrt{\frac{n!}{m!}}\frac{[f(m)]!}{[f(n)]!}\alpha^{m-n}L^{m-n}_{n}(\vert\alpha\vert^{2})\vert m\rangle,\;\;\;\;\;\;    m\geq n,
\end{array}
\right.
\end{eqnarray}
where $ [f(n)]! = f(n)f(n-1)...f(1) $ with the conventional relation $ [f(0)]! = 1 $, $L_{k}^{l}(x)=\sum_{r=0}^{\infty}\frac{(k+l)!}{(l+r)!(k-r)!}\frac{(-x)^{r}}{r!}$ corresponds to the associated Laguerre polynomials and $\mathcal{N}^{'}_{i}, i=1,2,$ refers to the normalization factors which are given by
\begin{eqnarray}\label{NDNS1-NF}
\mathcal{N}^{'}_{1} &=& \left(     \sum_{m=0}^{\infty} \frac{m!}{n!} \left( \frac{[f(m)]!}{[f(n)]!} \right)^{2}
\mathrm{e}^{- |\alpha|^{2}} |\alpha|^{2(n-m)} \left( L^{n-m}_{m}(\vert\alpha\vert^{2}) \right)^{2}     \right)^{-1/2},       \nonumber \\
\mathcal{N}^{'}_{2} &=& \left(     \sum_{m=0}^{\infty} \frac{n!}{m!} \left( \frac{[f(m)]!}{[f(n)]!} \right)^{2}
\mathrm{e}^{- |\alpha|^{2}} |\alpha|^{2(m-n)} \left( L^{m-n}_{m}(\vert\alpha\vert^{2}) \right)^{2}     \right)^{-1/2}.
\end{eqnarray}
Similarly, the exact form of the second type of the NDNSs reads as
\begin{eqnarray}\label{NDNS2-Expansion}
\hspace{-2cm}| \alpha ,f , n \rangle^{''} = \mathrm{e}^{- \frac{|\alpha|^{2}}{2}} \times
\left\{
\begin{array}{c}
\mathcal{N}^{''}_{1}\sum_{m=0}^{\infty}\sqrt{\frac{m!}{n!}}\frac{[f(n)]!}{[f(m)]!}(-\alpha^{*})^{n-m}L^{n-m}_{m}(\vert\alpha\vert^{2})   \vert m\rangle,\;\;\;\;\;\;       m\leq n, \\
\\
\mathcal{N}^{''}_{2}\sum_{m=0}^{\infty}\sqrt{\frac{n!}{m!}}\frac{[f(n)]!}{[f(m)]!}\alpha^{m-n}L^{m-n}_{n}(\vert\alpha\vert^{2})\vert m\rangle,\;\;\;\;\;\;    m\geq n,
\end{array}
\right.
\end{eqnarray}
with the following normalization constants
\begin{eqnarray}\label{NDNS2-NF}
\mathcal{N}^{''}_{1} &=& \left(     \sum_{m=0}^{\infty} \frac{m!}{n!} \left( \frac{[f(n)]!}{[f(m)]!} \right)^{2}
\mathrm{e}^{- |\alpha|^{2}} |\alpha|^{2(n-m)} \left( L^{n-m}_{m}(\vert\alpha\vert^{2}) \right)^{2}     \right)^{-1/2},       \nonumber \\
\mathcal{N}^{''}_{2} &=& \left(     \sum_{m=0}^{\infty} \frac{n!}{m!} \left( \frac{[f(n)]!}{[f(m)]!} \right)^{2}
\mathrm{e}^{- |\alpha|^{2}} |\alpha|^{2(m-n)} \left( L^{m-n}_{m}(\vert\alpha\vert^{2}) \right)^{2}     \right)^{-1/2}.
\end{eqnarray}
By looking deeply at the NDNSs obtained in (\ref{NDNS1-Expansion}) and (\ref{NDNS2-Expansion}) and comparing them with the introduced NDNSs in (\ref{NDNSs-Oliveira}), it is manifestly found that, they are essentially different from each other by the term $[f(n)]!$. We would like to emphasize the fact that the nonlinear terms $[f(n)]!$ and $[f(m)]!$ are logically obtained in our introduced state while the term $[f(n)]!$ which is appeared in  \cite{Oliveira2005} is not arisen from a reasonable procedure, since the authors have manually entered this term in DNSs. It is also valuable to state that based on our formalism, many NDNSs can be easily constructed by using various nonlinearity functions associated with nonlinear oscillators as well as every solvable quantum systems (due to the simple relation $e_{n} = n f^{2}(n)$ \cite{Roknizadeh.Tavassoly2004,Roknizadeh.Tavassoly2005}).
In the next section, by using the group theoretical method, another class of NDNS with particular nonlinearity function $f(n)$ is acquired.
 \section{Nonlinear displaced number states: Group theoretical approach}
It is illustrated that, by considering the group algebra and paying attention to the fact that the construction of a unitary displacement operator with $\hat{A}$ and $\hat{A}^{\dag}$ is possible through the particular nonlinearity functions associated with the specific physical systems, a few classes of nonlinear coherent states may be produced. Based on this fact, in the following, two types of NDNSs are introduced by using the group representation.
\begin{itemize}
\item \emph{Gilmore-Perelomov-type of $SU(1, 1)$ coherent states:}
Keeping in mind the approach of Man'ko {\it et al} in \cite{Manko.etal1997}, it is shown that, the (modified) trigonometric potential $ V(x)=U_{0} \tan^{2}(bx) $, in which $ U_{0}$ is the strength of the potential and $b$ is its range \cite{Nieto1978}, corresponds to the nonlinearity function $f_{GP}(n)$ which is given by \cite{Miry.Tavassoly2012}
\begin{eqnarray}\label{SU11Nonlinearity}
f_{GP}(n) = \sqrt{\frac{\hbar b^{2}}{2\mu \Omega}(n+2\lambda -1)}.
\end{eqnarray}
In the latter relation, $\Omega$ is the frequency of the field, $\mu$ is the mass of the particle, $\lambda$ is related to the potential strength and is sometimes the so-called Bargmann index, which can take any positive integers or half integers, i.e., $ \lambda=\frac{1}{2}, 1, \frac{3}{2}, 2,...$ . Also, the parameter $b$ denotes the potential range and is obtained via the relation $\lambda(\lambda + 1) = 2\mu U_{0} /  \hbar^{2}b^{2}$.
By substituting the nonlinearity function (\ref{SU11Nonlinearity}) into the $f$-deformed bosonic annihilation operator $ \hat{A}_{GP} = \hat{a} f_{GP}(\hat{n}) $, one may define the new operators
$ \hat{K}_{-} = \sqrt{\frac{2\mu \Omega}{\hbar b^{2}}} \hat{A}_{GP}, \hat{K}_{+} = \sqrt{\frac{2\mu \Omega}{\hbar b^{2}}}\hat{A}_{GP}^{\dag} $  and $ \hat{K}_{0} = \lambda + n $ satisfying the commutation relations $ [\hat{K}_{0},\hat{K}_{\pm}] = \pm \hat{K}_{\pm} $ and $ [\hat{K}_{-},\hat{K}_{+}] = 2 \hat{K}_{0} $, which are the well-known $su(1,1)$ Lie algebra \cite{Miry.Tavassoly2012}. Based on the group theoretical construction for the Gilmore-Perelomov approach corresponding to discrete series representation of $SU(1,1)$ group, the displacement operator reads as $ \hat{D}^{GP}_{f}(\alpha) = \exp \left( \xi \hat{K}_{+} - \xi^{*} \hat{K}_{-} \right) $ \cite{Gilmore1974,Perelomov1986}. Now, by the action of $\hat{D}^{GP}_{f}(\alpha)$ on the number state, the NDNSs associated with $SU(1,1)$ group are given by:
\begin{eqnarray}\label{NDNS-GP}
| \zeta ,f,n\rangle^{GP} &=& \hat{D}^{GP}_{f}(\alpha) | n \rangle   \nonumber \\
&=& (1-\vert \zeta\vert^{2})^{\lambda}\sum_{m=0}^{\infty} \sum_{p=0}^{\min[m,n]}
\frac{(-\zeta^{*})^{m}\zeta^{n}\sqrt{m!n!} \left(1-\frac{1}{\vert \zeta \vert^{2}}\right)^{p}}{p! (n-p)!(m-p)!} \nonumber \\
&\times& \frac{[f_{GP}(n)]![f_{GP}(m)!]}{[f_{GP}(p)]!^{2}} \vert m\rangle,
\end{eqnarray}
where we have used $\xi =\sqrt{\frac{\hbar b^{2}}{2\mu \Omega}}\alpha  $ and $\zeta=\frac{\xi}{\vert\xi\vert} \tanh \xi $ with $ |\xi|<1 $. The condition $|\zeta|<1$ implies the fact that the phase space of the $SU(1,1)$ coherent states is confined to the interior of the unit disk of the complex plane.
\item \emph{$SU(2)$ coherent states:}
As another case of a physical potential, which can be equivalent to a nonlinearity function, is known as the modified P{\"o}schl-Teller potential by relation $V(x) = U_{0} \tanh^{2}(ax)$ with $U_{0}$ and $a$ as the depth and the range of well, respectively \cite{Nieto1978}. This potential is related to a system that possesses a finite discrete spectrum. The corresponding nonlinearity function is of the form
\begin{eqnarray}\label{SU2Nonlinearity}
f_{SU(2)}(n)=\sqrt{\frac{\hbar a^{2}}{2\mu \Omega}(2 s + 1 - n )},
\end{eqnarray}
where $\mu$ denotes the reduced mass of the molecule and $s$ means the depth of well which is related to its range though the relation $s(s + 1) = 2 \mu U_{0}/\hbar^{2}a^{2}$.
Considering the $f$-deformed bosonic annihilation operator $ \hat{\mathcal{A}} = \hat{a} f_{SU(2)}(\hat{n}) $, the new operators,
$ \hat{\mathcal{K}}_{-} = \sqrt{\frac{2\mu \Omega}{\hbar a^{2}}} \hat{\mathcal{A}}, \hat{\mathcal{K}}_{+} = \sqrt{\frac{2\mu \Omega}{\hbar a^{2}}}\hat{\mathcal{A}}^{\dag} $  and $ \hat{\mathcal{K}}_{0} = n - s $ may be defined with the commutation relations $ [\hat{\mathcal{K}}_{0},\hat{\mathcal{K}}_{\pm}] = \pm \hat{\mathcal{K}}_{\pm} $ and $ [\hat{\mathcal{K}}_{-},\hat{\mathcal{K}}_{+}] = - 2 \hat{\mathcal{K}}_{0} $. Paying attention to the fact that the introduced operators clearly satisfy the $SU(2)$ Lie algebra \cite{Miry.Tavassoly2012}, the displacement-type operator corresponding to this group reads as $ \hat{D}^{SU(2)}_{f}(\alpha) = \exp \left( \eta \hat{\mathcal{K}}_{+} - \eta^{*} \hat{\mathcal{K}}_{-} \right)$ with $ \eta = \sqrt{\frac{\hbar a^{2}}{2\mu \Omega}} \alpha$. By the action of such a displacement operator on the number state, the new class of NDNSs associated with $SU(2)$ group is obtained by the following relation:
\begin{eqnarray}\label{NDNS-SU2}
| \gamma ,f,n\rangle^{SU(2)} &=& \hat{D}^{SU(2)}_{f}(\alpha) | n \rangle   \nonumber \\
&=& \left( \frac{1}{1 + | \gamma |^{2}} \right)^{s}\sum_{m=0}^{\infty}\sum_{p=0}^{\min[m,n]}
\frac{( - \gamma^{*})^{m} \gamma ^{n}\sqrt{m!n!}(\frac{-|\gamma|^{2}}{1+|\gamma|^{2}})^{(-p)}}{p! (n-p)! (m-p)!} \nonumber \\
&\times& \frac{[f_{SU(2)}(n)]![f_{SU(2)}(m)!]}{[f_{SU(2)}(p)]!^{2}} | m \rangle,
\end{eqnarray}
where $\gamma = \frac{\eta}{|\eta|} \tanh \eta$. It may be noted that the parameter $s$ can get values $\frac{1}{2}, 1, \frac{3}{2}, 2, ...$ . Adding our obtained results in two latter sections, it is seen that we have produced four different classes of NDNSs, which all of them have been introduced by some reasonable procedures.
Anyway, we are now in a position to examine the nonclassicality features of  the obtained NDNSs in the continuation of the paper.
\end{itemize}
 \section{Nonclassical criteria}
Since the nonclassical light is of special attention in the field of quantum optics and quantum information processing, in this section, we are going to study some of the well-known nonclassicality features of the introduced NDNSs. For this purpose, sub-Poissonian statistics as well as the negativity of Wigner distribution function are examined, numerically. Before going proceed, it ought to be mentioned that for evaluating any quantity for the NDNSs which have been produced by algebraic method (the relations (\ref{NDNS1-Expansion}) and (\ref{NDNS2-Expansion})), a nonlinearity function should be chosen. For this purpose, we use the nonlinearity function $ f(n) = ( 1 + k n)^{-1}$, which has been considered in \cite{Oliveira2005}.
%
 \subsection{Sub-Poissonian statistics: Mandel parameter}
%
This subsection deals with studying the quantum statistics of the states through the Mandel’s $Q$-parameter, which characterizes the photon statistics of light.  This parameter has been defined as \cite{Mandel1979}
\begin{eqnarray}\label{Mandel parameter}
Q_{M} = \frac{\langle (\hat{a}^{\dag}\hat{a})^{2}\rangle - \langle \hat{a}^{\dag} \hat{a}\rangle^{2}}{\langle \hat{a}^{\dag} \hat{a}\rangle}-1=\frac{\langle \hat{a}^{\dagger^{2}}\hat{a}^{2}\rangle - \langle \hat{a}^{\dagger}\hat{a} \rangle ^{2}}{\langle \hat{a}^{\dagger}\hat{a} \rangle}.
\end{eqnarray}
Whenever $-1\leq Q <0 \;(Q>0)$ the statistics is sub-Poissonian  (super-Poissonian) and $Q=0$ indicates the Poissonian statistics. By the way, the state vector of the system shows the nonclassical behavior when the photons statistics of field is sub-Poissonian. \\
Figure  \ref{Mandel-NDNSs} shows the Mandel parameter for some different classes of NDNSs corresponding to (a) $ | \alpha , f , n \rangle^{'} $, (b) $ | \alpha , f , n \rangle^{''} $, (c) $ | \zeta , f , n \rangle $ and (d) $ | \gamma , f , n \rangle $ corresponding to the relations (\ref{NDNS1-Expansion}), (\ref{NDNS2-Expansion}), (\ref{NDNS-GP}) and (\ref{NDNS-SU2}), respectively. It is seen from figure \ref{Mandel-NDNSs}(a) that, nonclassical behavior (sub-Poissonian statistics) is obviously observed in some intervals of $\alpha$. In addition, by increasing the value of $n$, this behaviour is strengthened. Unlike figure \ref{Mandel-NDNSs}(a), figure \ref{Mandel-NDNSs}(b) indicates that the maximum nonclassicality signs is occurred when $ n = 0 $ (nonlinear displaced {\it vacuum} state). In these two latter figures, it is seen that, by increasing $\alpha$, the  Mandel parameter gets negative values every where, i.e., the photon statistics of the field becomes full sub-Poissonian. Figures \ref{Mandel-NDNSs}(c) and (d) exhibit locally (around $\alpha = 0$) sub-Poissonian statistics, in which by an increase of $ \lambda $ or $s$, the space for which the photon statistics is sub-Poissonian, is gradually decreased.
%
 \subsection{Wigner distribution function}
%
The Wigner function, known as the earliest quasi-probability distribution function \cite{Wigner1932}, is a useful criterion which specifies the nonclassicality of the field. It is now valuable to declare that, although the Wigner function, in the sense that it is a distribution function, will have to be commonly positive, but there may exist some finite regions in phase space of the Wigner function of  a quantum state, in which this function gets negative values; the fact that is called `nonclassicality feature'. The Wigner function associated with any quantum state can be expressed as \cite{Gerry.Knight2005,Vogel.Welsch2006}
\begin{eqnarray}\label{WignerDef}
W(\alpha ,\alpha^{*})=\frac{2}{\pi}\sum_{n = 0}^{\infty}(-1)^{n}\langle n, \alpha | \hat{\rho} | n, \alpha \rangle,
\end{eqnarray}
where $ | n, \alpha \rangle = \hat{D}(\alpha) | n \rangle $ is the displaced number state introduced in \cite{Oliveira1990} and $\hat{\rho}$ denotes the density matrix operator of quantum state. Considering the obtained NDNSs via algebraic method in (\ref{NDNS1-Expansion}) and (\ref{NDNS2-Expansion}), the corresponding Wigner functions may be evaluated in the form:
\begin{eqnarray}\label{NDNS1-Wigner}
\hspace{-2cm}W^{'}(\alpha,\alpha^{*}) = \frac{2 \; \mathrm{e}^{-(|\alpha|^{2}+|\beta|^{2})}}{\pi [f(n)]!^{2}} \times
\left\{
\begin{array}{c}
\bigg|\mathcal{N}^{'}_{1}\sum_{m = 0}^{\infty}\sqrt{\frac{k!}{n!}}(-\alpha^{*})^{n-m}\beta^{m-k}[f(m)]!  \\
\times L^{n-m}_{m}(|\alpha|^{2})L^{m-k}_{k}(|\beta|^{2})\bigg|^{2},   \hspace{1cm}    m\leq n,    \\
\\
\bigg|\mathcal{N}'_{2}\sum_{m = 0}^{\infty} \frac{\sqrt{n!k!}}{m!}\alpha^{n-m}\beta^{m-k}[f(m)]! \\
\times L^{m-n}_{n}(|\alpha|^{2})L^{m-k}_{k}(|\beta|^{2})\bigg|^{2} ,   \hspace{1cm}    m\geq n,
\end{array}
\right.
\end{eqnarray}
\begin{eqnarray}\label{NDNS2-Wigner}
\hspace{-2cm}W^{''}(\alpha,\alpha^{*}) = \frac{2[f(n)]!^{2} \mathrm{e}^{-(|\alpha|^{2}+|\beta|^{2})}}{\pi } \times
\left\{
\begin{array}{c}
\bigg|\mathcal{N}^{''}_{1}\sum_{m = 0}^{\infty}\sqrt{\frac{k!}{n!}}\frac{(-\alpha^{*})^{n-m}}{[f(m)]!}\beta^{m-k} \\
\times L^{n-m}_{m}(|\alpha|^{2})L^{m-k}_{k}(|\beta|^{2})\bigg|^{2},   \hspace{1cm}   m\leq n,   \\
\\
\bigg|\mathcal{N}^{''}_{2}\sum_{m = 0}^{\infty}\frac{\sqrt{n!k!}}{m!}\frac{\alpha^{n-m}}{[f(m)]!}\beta^{m-k}  \\
\times L^{m-n}_{n}(|\alpha|^{2})L^{m-k}_{k}(|\beta|^{2})\bigg|^{2},   \hspace{1cm}     m\geq n.
\end{array}
\right.
\end{eqnarray}
Similarly, the Wigner function associated with NDNSs of $SU(1,1)$ and $SU(2)$ groups are respectively evaluated as below:
\begin{eqnarray}\label{WignerSU11}
W_{\mathrm{GP}}(\alpha,\alpha^{*}) &=& \frac{2 \; \mathrm{e}^{-|\alpha|^{2}}(1-|\zeta|^{2})^{2\lambda}}{\pi}\sum_{k = 0}^{+\infty}(-1)^{k}  \nonumber \\
&\times& \Bigg|\sum_{m = 0}^{k}\sum_{p = 0}^{\min[m,n]}\sqrt{\frac{n!}{k!}}
 \frac{m!(-\alpha)^{k-m}L_{m}^{k-m}(|\alpha|^{2})\zeta^{m}(-\zeta^{*})^{n}(1-|\zeta|^{-2})^{p}}{p!(m-p)!(n-p)!} \nonumber \\
 &\times&  \frac{[f_{\mathrm{GP}}(n)]![f_{\mathrm{GP}}(m)]!}{[f_{\mathrm{GP}}(p)]!^{2}}  \nonumber \\
 &+& \sum_{m=k+1}^{+\infty}\sum_{p = 0}^{\min[m,n]}\sqrt{k!n!}
 \frac{(\alpha)^{m-k}L_{k}^{m-k}(|\alpha|^{2})\zeta^{m}(-\zeta^{*})^{n}(1-|\zeta|^{-2})^{p}}{p!(m-p)!(n-p)!}   \nonumber \\
 &\times&  \frac{[f_{\mathrm{GP}}(n)]![f_{\mathrm{GP}}(m)]!}{[f_{\mathrm{GP}}(p)]!^{2}}\Bigg|^{2},
\end{eqnarray}
\begin{eqnarray}\label{WignerSU2}
W_{SU(2)}(\alpha,\alpha^{*}) &=& \frac{2 \; \mathrm{e}^{-|\alpha|^{2}}(1+|\gamma|^{2})^{- 2s}}{\pi}\sum_{k = 0}^{+\infty}(-1)^{k} \nonumber \\
&\times& \Bigg|\sum_{m = 0}^{k}\sum_{p = 0}^{\min[m,n]}\sqrt{\frac{n!}{k!}}
 \frac{m!(-\alpha)^{k-m} L_{m}^{k-m}(|\alpha|^{2})\gamma^{m}(-\gamma^{*})^{n}(\frac{-|\gamma|^{2}}{1+|\gamma|^{2}})^{-p}}{p!(m-p)!(n-p)!}  \nonumber \\
 &\times& \frac{[f_{SU(2)}(n)]![f_{SU(2)}(m)]!}{[f_{SU(2)}(p)]!^{2}} \nonumber \\
 &+& \sum_{m=k+1}^{+\infty}\sum_{p = 0}^{\min[m,n]}\sqrt{k!n!}
 \frac{(\alpha)^{m-k}L_{k}^{m-k}(|\alpha|^{2})\gamma^{m}(-\gamma^{*})^{n}(\frac{-|\gamma|^{2}}{1+|\gamma|^{2}})^{-p}}{p!(m-p)!(n-p)!} \nonumber \\
&\times& \frac{[f_{SU(2)}(n)]![f_{SU(2)}(m)]!}{[f_{SU(2)}(p)]!^{2}} \Bigg|^{2}.
\end{eqnarray}
In figure \ref{Wigner-NDNSs}, we have plotted the Wigner distribution function of the NDNSs obtained in relations (\ref{NDNS1-Expansion}), (\ref{NDNS2-Expansion}), (\ref{NDNS-GP}) and (\ref{NDNS-SU2}) for the same chosen parameters mentioned in figure \ref{Mandel-NDNSs}. Figures \ref{Wigner-NDNSs}(a)--(d) indicate clearly the negativity of Wigner function in some finite regions of phase space, which implies the fact that, the introduced NDNSs are `nonclassical'. It is also valuable to state that, by comparing quantitatively figure \ref{Wigner-NDNSs}(a) with figures \ref{Wigner-NDNSs}(b)--(d), it is seen that the amount of the negativity of Wigner function (the depth of this nonclassicality feature) in figure \ref{Wigner-NDNSs}(a) is nearly $10$ times greater than the others. In other words, the strength of nonclassicality of the state in (\ref{NDNS1-Expansion}) is more visible than the other states in (\ref{NDNS2-Expansion}), (\ref{NDNS-GP}) and (\ref{NDNS-SU2}).
%
 \section{Summary and conclusion}\label{Summary}
In this paper, by modifying the formalism of NDNSs presented in \cite{Oliveira2005}, we have introduced four distinct classes of NDNSs through algebraic and group treatments. For this purpose, by considering the DNSs together with nonlinear coherent states approach, two distinct classes of NDNSs were reasonably obtained via an algebraic treatment. In addition, by using a special class of Gilmore-Perelomov-type of $SU(1,1)$ and a class of $SU(2)$ coherent states (group approach), two other NDNSs were also introduced. Then, in order to study the nonclassicality features of the introduced states, sub-Poissonian statistics by evaluating Mandel parameter and the variation of Wigner quasi-probability distribution function associated with the obtained NDNSs were numerically examined. The presented results showed that the NDNSs exhibit sub-Poissonian statics (nonclassical behaviour) in a finite region. Also, as another appearance of the nonclassicality signs of the NDNSs, it was observed that the Wigner function is also negative in some areas of phase space. This means that the NDNSs can be considered as a good candidate for nonclassical light.



\providecommand{\newblock}{}

 \end{document}